\documentclass[twocolumn,showpacs,superscriptaddress,twoside,prl]{revtex4-1}
\usepackage{amsmath,amssymb,url,graphicx,epstopdf,bm}
\usepackage{braket}
\usepackage{epsfig,hyperref,amsmath,amssymb,amsfonts, latexsym, dsfont}
\usepackage{wasysym,multirow,mathrsfs,xcolor,color}
\usepackage[caption=false]{subfig}

\begin{document}

\title{Unity-Efficiency Parametric Down-Conversion \\
	via Amplitude Amplification }
\author{Murphy Yuezhen Niu}
\email{yzniu@mit.edu}
\affiliation{Research Laboratory of Electronics, Massachusetts Institute of Technology, Cambridge, Massachusetts 02139, USA}
\affiliation{Department of Physics, Massachusetts Institute of Technology, Cambridge, Massachusetts 02139, USA}
\author{Barry C. Sanders}
\affiliation{Institute for Quantum Science and Technology, University of Calgary, Alberta, Canada T2N 1N4}
\affiliation{%
	Program in Quantum Information Science,
	Canadian Institute for Advanced Research, Toronto, Ontario M5G 1Z8, Canada}
\affiliation{%
	Hefei National Laboratory for Physical Sciences at the Microscale,
	University of Science and Technology of China, Anhui 230026, China}
\affiliation{%
	Shanghai Branch, CAS Center for Excellence and Synergetic Innovation Center in Quantum Information and Quantum Physics, University of Science and Technology of China, Shanghai 201315, China}
	\affiliation{%
	Institute for Quantum Information and Matter, California Institute of Technology, Pasadena, California 91125, USA}
\author{Franco N. C. Wong}
\affiliation{Research Laboratory of Electronics, Massachusetts Institute of Technology, Cambridge, Massachusetts 02139, USA}
\author{Jeffrey H. Shapiro}
\affiliation{Research Laboratory of Electronics, Massachusetts Institute of Technology, Cambridge, Massachusetts 02139, USA}
\begin{abstract}
We propose an optical scheme, employing optical parametric down-converters interlaced with nonlinear sign gates (NSGs), that completely converts an $n$-photon Fock-state pump to $n$ signal-idler photon pairs when the down-converters' crystal lengths are chosen appropriately.  The proof of this assertion relies on amplitude amplification, analogous to that employed in Grover search, applied to the full quantum dynamics of single-mode parametric down-conversion.  When we require that all Grover iterations use the same crystal, and account for potential experimental limitations on crystal-length precision, our optimized conversion efficiencies reach unity for $1\le n \le 5$, after which they decrease monotonically for $n$ values up to 50, which is the upper limit of our numerical dynamics evaluations.  Nevertheless, our conversion efficiencies remain higher than those for a conventional (no NSGs) down-converter. 
\end{abstract}

\date{\today}
\maketitle

Nonclassical states of light, such as single-photon states~\cite{single1, single2, single3}, polarization-entangled states~\cite{entangled1, entangled2} and multi-photon path-entangled states~\cite{multi1,multi2, multi3, multi4} are essential for linear-optical quantum computation~\cite{Knill2001}, quantum  communication~\cite{BB84,Ekert,CV}, quantum metrology~\cite{Holland1993,Sanders1995}, and experimental tests of quantum foundations \cite{CHSH, Bell3, Bell4}. Spontaneous parametric down-conversion~(SPDC) employing the  $\chi^{(2)}$ nonlinearity~\cite{entangled1} is a standard tool for generating nonclassical light.   As currently implemented, SPDC sources of nonclassical light rely on strong coherent-state pump beams.  These pumps do not suffer appreciable depletion in the down-conversion process, meaning that their conversion efficiencies are exceedingly low.  Moreover, the number of signal-idler pairs that are emitted in response to a pump pulse is random.    To circumvent these drawbacks, we focus our attention on SPDC using $n$-photon Fock-state pumps~\cite{Motes2016}.  
We propose and analyze a scheme using such pumps that interlaces SPDC processes with nonlinear sign gates (NSGs)~\cite{Knill2001} to generate $n$ signal-idler pairs with unity efficiency when the down-converters' crystal lengths are chosen appropriately.  Our proof of unity-efficiency conversion presumes $n\gg 1$ and allows each Grover iteration to employ a different crystal length.  Because the precision with which those crystal lengths must be realized  becomes increasingly demanding as $n$ increases, we evaluate the conversion efficiencies at a fixed crystal-length precision.  Furthermore, to reduce our scheme's resource burden, we perform our efficiency evaluations assuming that all Grover iterations use the same crystal.  We find that complete conversion is maintained for $1 \le n \le 5$, and that our approach's conversion efficiencies---although less than 100\%---still exceed those of a conventional (no NSGs) down-converter for $n$ values up to 50.   Thus, even using the same crystal for all Grover iterations with finite crystal-length precision, our approach can efficiently prepare heralded single-photon states as well as dual-Fock ($|n\rangle|n\rangle$) states and multi-photon path-entangled states for $n\le5$~\cite{inprep}.
     
We begin by solving the full quantum dynamics for SPDC with single-mode signal, idler, and pump beams.  Conventionally, SPDC dynamics are derived under the nondepleting-pump assumption, which treats a strong coherent-state pump as a constant-strength classical field throughout the nonlinear interaction. To date, SPDC with a quantized pump field~\cite{Rubin,Diana} has only been solved for pump-photon numbers up to 4~\cite{Kim}. We construct the SPDC solution for an arbitrary single-mode pure-state pump as an iteration that we can evaluate numerically for pump photon numbers up to 50. From this result, we prove a fundamental bound on SPDC's conversion efficiency: no pure-state pump whose average photon number exceeds one can be completely converted to signal-idler photon pairs.

Inspired by the Grover search algorithm's use of amplitude amplification~\cite{Lov1996,Brassard2002}, we show how the preceding limit on SPDC's conversion efficiency can be transcended by employing NSGs in between SPDC processes.  In particular, we show that our method increases the efficiency with which all pump photons are converted to signal-idler pairs, enabling complete pump conversion to be achieved for Fock-state pumps when the down-converters' crystal lengths are chosen appropriately.  This perfect conversion is deterministic if the NSGs are implemented using nonlinear optical elements.  It is postselected---based on ancilla-photon detections---if the NSGs are realized with only linear optics.  

Our technique for unity-efficiency parametric down-conversion (UPDC) has transformative applications in quantum metrology, quantum cryptography and quantum computation. In quantum metrology, an interferometer whose two input ports are illuminated by the signal and idler of the $n$-pair (dual-Fock) state~$\ket{n,n}$ achieves a quadratic improvement in phase-sensing accuracy over what results from sending all $2n$ photons into one input port~\cite{Holland1993}.  Single-mode SPDC yields a thermal distribution of~$\ket{n,n}$ states, however, which erases the preceding entanglement-based advantage~\cite{Sanders1995}, whereas UPDC delivers the desired dual-Fock state for this purpose~(Sec.~II of \cite{groverAppendix}). The dual-Fock state turns out to be extremely valuable for preparing heralded Greenberger-Horne-Zeilinger (GHZ) and other path-entangled states with high probability, which are crucial resources for device-independent quantum cryptography~\cite{Gisin, Gisin2}, quantum secret sharing~\cite{Hiller}, and testing quantum nonlocality~\cite{Pan2000}.

Our development begins by addressing the $t \ge 0$ quantum dynamics for parametric down-conversion with single-mode signal, idler, and pump beams.  The relevant three-wave-mixing interaction Hamiltonian is~\cite{Rubin}
\begin{align}
	\hat{H}
		=\text{i}\hbar \kappa\!\left(\hat{a}_s^{\dagger}\hat{a}_i^{\dagger}\hat{a}_p - \hat{a}_p^{\dagger}\hat{a}_s\hat{a}_i\right),
\label{HSPDC}
\end{align}
where~$\hat{a}_j^{\dagger}$ ($\hat{a}_j$) is the photon creation (annihilation) operator
and $j= s,i,p$ denotes the signal, idler, and pump, respectively.
The coefficient $\kappa$, which is assumed to be real valued,  characterizes the nonlinear susceptibility $\chi^{(2)} $ of the down-conversion crystal~\cite{Rubin}. We assume SPDC with type-II phase matching, so that the signal and idler beams are orthogonally polarized and the pump is co-polarized with the idler.  This orthogonality is crucial to realizing the Grover iteration, as detailed below.

We restrict ourselves to initial states of the form 
$\ket{\Psi(0)} = \sum_{n=0}^\infty c_n\ket{\Psi_n(0)}$,
where $\sum_{n=0}^\infty |c_n|^2 = 1$, and 
\begin{equation}
\ket{\Psi_n(0)} = \sum_{k=0}^n f_k^{(n)}(0) \ket{k,k,n-k},
\label{basisSPDCinitial}
\end{equation}
with $\sum_{k=0}^n |f_k^{(n)}(0)|^2 = 1$, and $\ket{n_s,n_i,n_p}$ being the Fock state containing $n_s$ signal photons, $n_i$ idler photons, and $n_p$ pump photons.  For these initial states, the SPDC dynamics occur independently in the subspaces spanned by $\{\ket{0, 0,n},\ket{1,1,n-1} ,\dots ,  \ket{n,n,0} : 0\le n < \infty\}$, whose basis states comprise all possibilities from no conversion to complete conversion of pump photons into signal-idler photon pairs.   The decoupling between these $n$-pump-photon subspaces allows us to solve the Schr\"{o}dinger equation, $\text{i}\hbar\ket{\dot{\Psi}(t)}=\hat{H} \ket{\Psi(t)}$ for $t\ge 0$, by solving the coupled ordinary differential equations 
\begin{widetext}
\begin{align}
\dot{f}_k^{(n)}(t)=
		\begin{cases}
			-\kappa\sqrt{n}f_{1}^{(n)}(t), & k=0\\
			\kappa\left[k\sqrt{n-k+1}f_{k-1}^{(n)}(t)- (k+1)\sqrt{n-k}f_{k+1}^{(n)}(t)\right], & k=1,2,\ldots, n-1\\
			\kappa nf_{n-1}^{(n)}(t), & k=n,
		\end{cases}
\label{ode}
\end{align}
\end{widetext}
given the initial conditions $\{f_k^{(n)}(0) : 0\le k \le n\}$. We then get the $n$-pump-photon subspace's state evolution,
\begin{equation}
	\ket{\Psi_n(t)}
		=\sum_{k=0}^n f_k^{(n)}(t) \ket{k,k,n-k},
\label{basisSPDC}
\end{equation}
from which the full state evolution, 
\begin{equation}
\ket{\Psi(t)} =\sum_{n=0}^\infty c_n \ket{\Psi_n(t)},
\label{fullstate}
\end{equation} 
follows. 
We have obtained analytical solutions to Eqs.~\eqref{ode} for $0 \le n\leq 4$,  and numerical solutions for $5\le n \le 50$.
The $n$th subspace's quantum conversion efficiency,
\begin{equation}
\mu_n(t) \equiv \sum_{k=1}^n \frac{k|f_k^{(n)}(t)|^2}{n}, \mbox{ when $\ket{\Psi_n(0)} = \ket{0,0,n}$},  
\end{equation}
is the fraction of the initial $n$ pump photons that are converted to signal-idler photon pairs.  The down-converter's total quantum conversion efficiency is then
\begin{equation}
\mu(t) \equiv \frac{\sum_{n=0}^{\infty}|c_n|^2 n \mu_n(t)}{\sum_{n=0}^{\infty}|c_n|^2 n}.
\end{equation}
Because $\sum_{k=0}^n|f_k^{(n)}(t)|^2= 1$ for all $n$, neither $\mu_n(t)$ nor $\mu(t)$ can exceed unity.  The central question for this paper is how to obtain unity-efficiency conversion, which occurs for $\mu_n(t)$ when $|f_n^{(n)}(t)| = 1$, and for $\mu(t)$ when $|f_n^{(n)}(t)| = 1$  for all $n$ with nonzero $c_n$.  

Our analytic solutions to Eqs.~\eqref{ode} for $1\le n \le 4$ with $\ket{\Psi_n(0)} = \ket{0,0,n}$ show that $\max_t[\mu_n(t)]$ decreases with increasing $n$ from $\max_t[\mu_1(t)] = 1$.  This downward trend in conversion efficiency continues for $5\le n \le 50$, where we employed numerical solutions because the Abel-Ruffini theorem shows that polynomial equations of fifth or higher order do not have universal analytic solutions.  In other words, when the down-converter crystal is driven by vacuum signal and idler and an $n$-photon Fock-state pump, \emph{only} the $n=1$ case can yield unity efficiency.  Moreover, because mixed states are convex combinations of pure states, exciting the down-converter with a mixture of $\ket{0,0,n}$ states also fails to realize complete conversion of pump photons to signal-idler photon pairs.  

To overcome this fundamental limitation we interlace SPDC processes with NSGs.  In Grover search~\cite{Lov1996}, NSGs serve as  quantum oracles that flip the sign of the marked state $\ket{n}$ by means of the unitary transformation
\begin{align}
	U_\text{NSG}^{(n)}\sum_{j=0}^n\alpha_j\ket{j}
		=\sum_{j=0}^n(-1)^{\delta_{jn}}\alpha_j\ket{j},
\end{align}
where $\delta_{jn}$ is the Kronecker delta function.  The $U_{\rm NSG}^{(2)}$ gate, which is essential to linear-optical quantum computing's construction of a CNOT gate~\cite{Knill2001}, has a nondeterministic implementation that only requires linear optics and single-photon detection. A deterministic realization of $U_{\rm NSG}^{(2)}$ is possible through use of a Kerr nonlinearity~\cite{Kerr}. Nondeterministic $U_{\rm NSG}^{(n)}$ gates have postselection success probabilities with $O(1/n^2)$ scaling~\cite{Scheel2005}. 

Grover search~\cite{Lov1996} finds the marked item in an unsorted data set of size $N$ in the optimal~\cite{Ikebook} $O(\sqrt{N})$ steps, as opposed to the best classical algorithm's requirement of $O(N)$ steps.  To reap Grover search's benefit in our context we perform it in the Fock basis. In particular, given a Fock-state input $\ket{0,0,n}$, with $n \ge 2$, our UPDC procedure uses $O(\sqrt{n})$ iterations of Grover search---in which an iteration consists of an NSG followed by SPDC---to convert that input to the dual-Fock-state output $\ket{n,n,0}$ with unity efficiency for $n$ sufficiently large.  (In Sec.~I of \cite{groverAppendix} we show that unity-efficiency conversion of $\ket{0,0,1}$ to $\ket{1,1,0}$ can be realized with a single SPDC stage.)  Our UPDC procedure is as follows.
\begin{enumerate}
\item[I.] \textit{Initialization: }
Initialize the UPDC procedure by sending signal, idler, and pump inputs in the joint state $\ket{0,0,n}$ into a length-$L_0$, type-II phase-matched $\chi^{(2)}$ crystal for an interaction time $t_0 = L_0/v$, where $v$ is the \emph{in situ} propagation velocity, to obtain the initial state~\cite{Lidar1999} 
\begin{align}
\ket{\Psi_0}=& \sum_{k=0}^n f_k^{(n,0)}(t_0)\ket{k, k,n-k},\label{stepI}
\end{align}
where the $\{f_k^{(n,0)}(t_0)\}$ are solutions to~(\ref{ode}) for the initial conditions $f_k^{(n,0)}(0)= \delta_{k0}$.
\item[II.] \textit{Sign flip on the marked state:} \label{phase shift}
Begin the $m$th Grover iteration by sending the signal, idler, and pump outputs from the $(m-1)$th iteration---whose joint state is
\begin{equation}
\ket{\Psi'_{m-1}} = \sum_{k=0}^n f_k^{(n,m-1)}(t_{m-1})\ket{k, k,n-k}, \mbox{ for $m \ge 1$,}
\label{mInPhi}
\end{equation} 
where $\ket{\Psi'_0} \equiv \ket{\Psi_0}$---through a polarization beam splitter~(PBS) to separate the signal and idler into distinct spatial modes with the pump accompanying the idler.   Then apply the $U_{\rm NSG}^{(n)}$ gate to the signal mode in Eq.~(\ref{mInPhi}) to produce the state
\begin{align}\label{phi}
\ket{\Psi_m}&= \sum_{k=0}^n f_k^{(n,m)}(0)\ket{k, k,n-k},
\end{align}
where $f_k^{(n,m)}(0) = (-1)^{\delta_{kn}}f_k^{(n,m-1)}(t_{m-1})$,
and use another PBS to recombine the signal, idler, and pump into a common spatial mode without changing their joint state.
\item[III.] \textit{Rotation toward the marked state:}  
Complete the $m$th Grover iteration by sending the signal, idler, and pump in the joint state $|\Psi_m\rangle$ into a length-$L_m$, type-II phase-matched $\chi^{(2)}$ crystal for an interaction time $t_m= L_m/v$ to obtain the state
\begin{align}
\ket{\Psi'_m}=& \sum_{k=0}^n f_k^{(n,m)}(t_m)\ket{k, k,n-k},\label{stepIII}
\end{align}
where the $\{f_k^{(n,m)}(t_m)\}$ are solutions to~(\ref{ode}) for the initial conditions $\{f_k^{(n,m)}(0)\}$.
\item[IV.] \textit{Termination:} Repeat Steps~II and III until the probability that Step~III's output beams are in the desired fully converted state is sufficiently close to unity.
\end{enumerate}
Below we explain how Steps~I--III can drive the conversion efficiency arbitrarily close to unity, and how, for $n$ sufficiently large, this can be done in $O(\sqrt{n})$ Grover iterations. 

For an initial state $|0,0,n\rangle$, the Fock-state amplitudes occurring in our UPDC procedure are real valued.  Thus, for our present purposes, we can reduce the UPDC procedure's state evolution to SU(2) rotations by writing 
\begin{equation}
|\Psi_m'\rangle = \sqrt{1-[f_n^{(n,m)}(t_m)]^2}\,|0\rangle + f_n^{(n,m)}(t_m)|1\rangle, \label{SU2states}
\end{equation}
for $m\ge 0$, where $|1\rangle \equiv |n,n,0\rangle$ is the fully converted state, and $|0\rangle$ is the $m$-dependent, normalized state satisfying $\langle 1|0\rangle = 0$.  In Sec.~I of~\cite{groverAppendix} we show that with $L_0$ appropriately chosen we can realize
\begin{equation}
\ket{\Psi_0'}  = \cos(\theta_g/2)\ket{0} + \sin(\theta_g/2)\ket{1},
\end{equation}
for small values of $\theta_g$; e.g., $\theta_g \simeq 1/\sqrt{n}$ for large $n$.  There we also prove that our UPDC procedure, with the $\{L_m\}$ appropriately chosen, can produce
\begin{equation}
\ket{\Psi_m'}  = \cos[(2m+1)\theta_g/2]\ket{0} + \sin[(2m+1)\theta_g/2]\ket{1},
\end{equation}
for $m>1$.  Terminating the UPDC procedure after $M$ Grover iterations, where $M$ is the largest integer satisfying $(2M+1)\theta_g \le \pi$, then gives a $\sin^2[(2M+1)\theta_g/2]$ conversion efficiency.  Rewriting this conversion efficiency as $1-\cos^2[(\pi-(2M+1)\theta_g)/2]$ and choosing $L_0$ such that $0 < (\pi-(2M+1)\theta_g)/2 \ll 1$, we find that $1-\cos^2[(\pi-(2M+1)\theta_g)/2] \approx1- (\pi-(2M+1)\theta_g)^2/4 \approx 1$. Moreover, for $\theta_g\simeq 1/\sqrt{n}$ with $n\gg 1$, we have that this near-unity conversion efficiency is realized with $M$ being $O(\sqrt{n})$, meaning that $\sqrt{n}$ iterations suffice to achieve that performance. 

Our proof that UPDC can achieve unity-efficiency conversion of an initial $|0,0,n\rangle$ state to a final $|n,n,0\rangle$ state 
for $n\gg 1$ allows each Grover iteration to use a crystal of a different length, making its required resources of order $O(\sqrt{n})$.  Thus in our analytic (for $2\le n \le 4$) and numerical (for $5\le n \le 50$) conversion-efficiency evaluations we restricted our procedure's Grover iterations to recirculate the signal, idler, and pump beams through a single length-$L_1$ crystal, and we chose $L_0$ and $L_1$ to maximize the conversion efficiency.  However, as Eqs.~\eqref{ode} evolutions have eigenmodes with associated eigenvalues whose magnitudes grow with increasing $n$, the precision to which the crystal lengths $L_0$ and $L_1$ must be cut grows with increasing $n$.  Thus, for experimental feasibility, our conversion-efficiency optimizations took $L_0$ and $L_1$ to be integer multiples of $10^{-3}v/\kappa$   ~\cite{footnote}.

\begin{figure}
\includegraphics[width=1.0\linewidth]{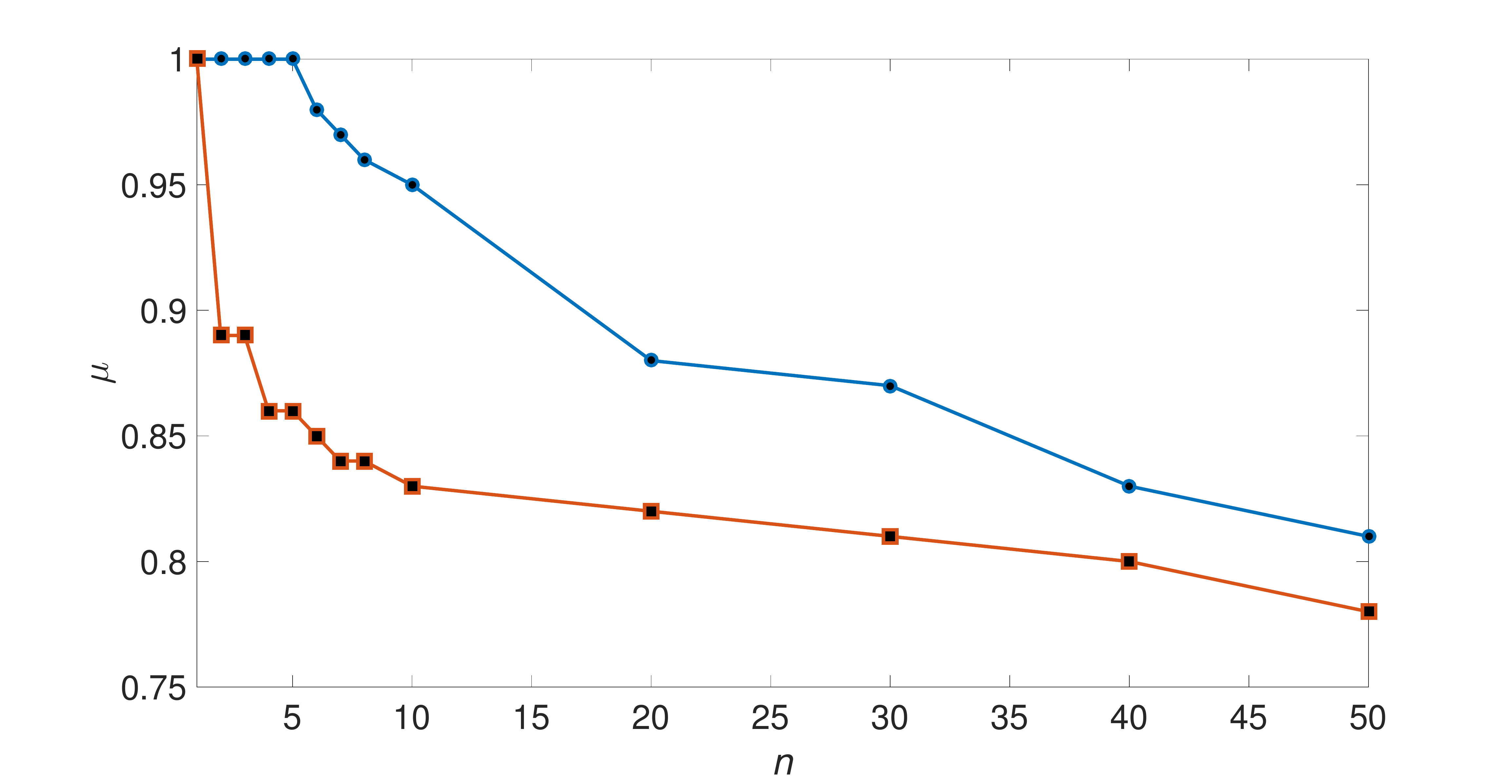}
\caption{Down-conversion efficiencies for $n$-photon Fock-state pumps optimized over nonlinear-crystal lengths cut to a precision of $10^{-3}v/\kappa$.  Lower (red) curve:  maximum conversion efficiencies for a $\chi^{(2)}$ crystal without Grover-search amplitude amplification.  Upper (blue) curve:  maximum UPDC conversion efficiencies, where  the $n=1$ point did not employ an NSG.}\label{efficiency}
\end{figure}

Available analytic solutions to Eqs.~\eqref{ode} for $n \le 4$ allowed us to verify that unity-efficiency conversion can be achieved for those pump-photon numbers; see Sec.~III of \cite{groverAppendix} for a demonstration that a single Grover iteration suffices for $n=2$.  For $n\in \{2,3,4,5,6,7,8,10,20,40,50\}$ the optimized conversion efficiencies we obtained are shown in Fig.~\ref{efficiency}.  Here we see that unity-efficiency conversion is possible for $n$ values up to 5, using a single Grover-iteration crystal that is cut with the assumed length precision.  Beyond $n=5$, however, greater precision is presumably required.  Figure~\ref{efficiency} also includes similarly evaluated conversion efficiencies for a conventional SPDC setup, i.e., one in which a single nonlinear crystal is employed without any NSGs.  As mentioned earlier, the conventional approach can only reach unity-efficiency conversion for $n=1$, and Fig.~\ref{efficiency} shows that the UPDC approach with finite crystal precision outperforms the conventional setup with the same crystal precision for $2\le n \le 50$.   Our UPDC conversion efficiencies presume the use of deterministic (unity efficiency) NSGs, such as can be realized under ideal conditions with a weak Kerr nonlinearity~\cite{Kerr} or with trapped atoms governed by the Jaynes-Cummings Hamiltonian~\cite{Azuma2011}. Now consider a UPDC procedure that employs $\sqrt{n}$ Grover iterations to transform an $n$-pump-photon Fock state to $n$ signal-idler photon pairs using nondeterministic NSGs.  Its conversion efficiency is reduced from our deterministic NSG result by a $(1/n^2)^{\sqrt{n}}$ factor, owing to each of its $\sqrt{n}$ NSG uses having an efficiency that is bounded above by $1/n^2$~\cite{Scheel2005}. Furthermore, each of these nondeterministic NSGs will require at least $n$ single-photon ancillae~\cite{Scheel2005}. 

\begin{figure}
\includegraphics[width=1.00\linewidth]{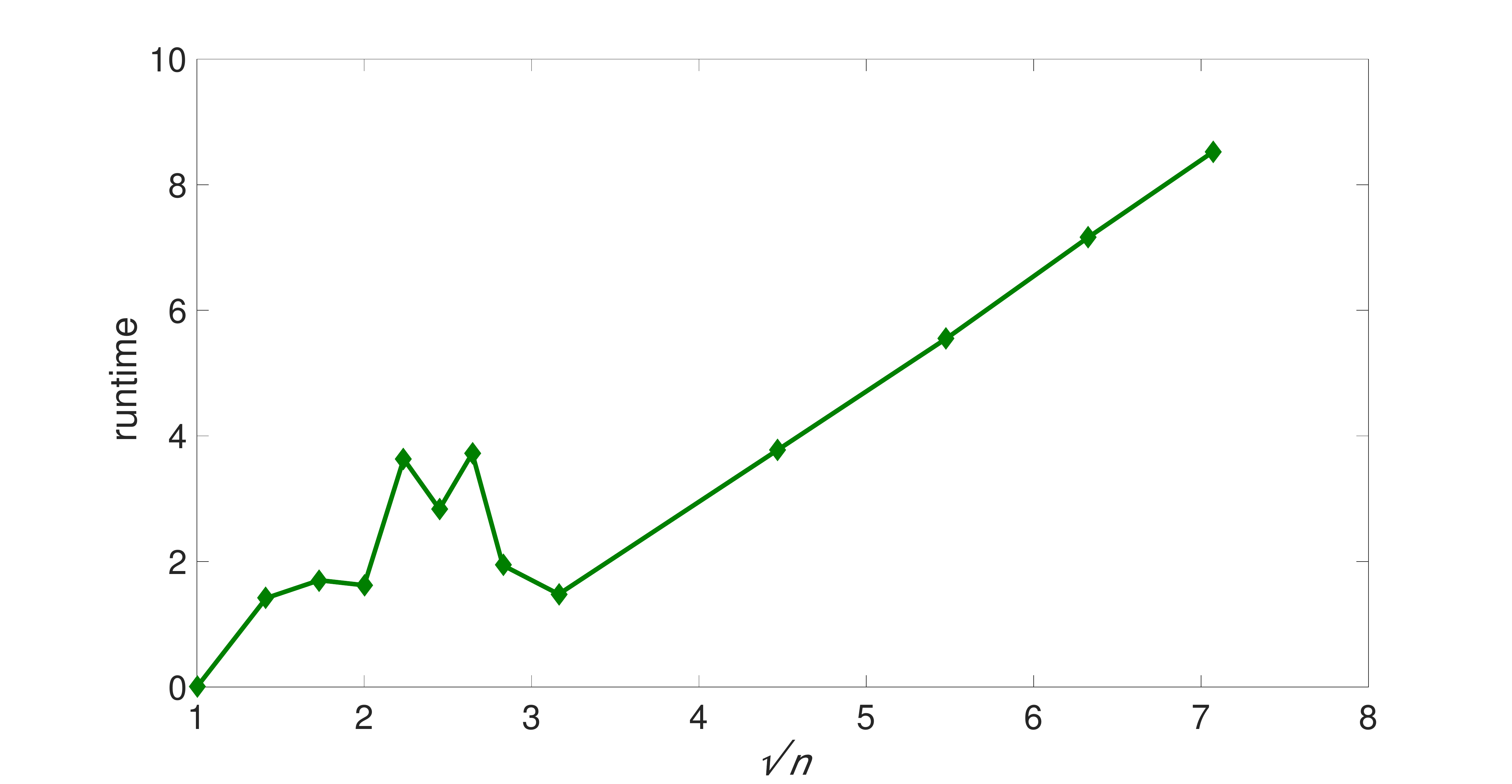}
\caption{UPDC runtime (defined to be $M_nL_1/v$ with $M_n$ being the number of Grover iterations used in Fig.~\ref{efficiency} to achieve maximum efficiency for an $n$-photon pump) versus $\sqrt{n}$.}\label{runtime}
\end{figure}

The preceding efficiency optimization also permits us to determine the runtimes for our UPDC procedure at finite crystal-length precision, where runtime is defined to be $M_nL_1/v$ with $M_n$ being the number of Grover iterations needed to achieve the $n$-photon-pump's maximum efficiency from Fig.~\ref{efficiency}.  These runtimes, which we have plotted in Fig.~\ref{runtime}, show the expected $O(\sqrt{n})$ behavior for $3 \le \sqrt{n} \le 7$.  

At this juncture, some discussion of implementation considerations is warranted. UPDC requires a very strong $\chi^{(2)} $ nonlinearity if it is to be practical.  Probably the most promising candidate for implementation is the induced $\chi^{(2)}$ behavior of the $\chi^{(3)}$ nonlinearity which uses nondegenerate four-wave mixing with a strong, nondepleting pump beam at one wavelength whose presence induces a strong $\chi^{(2)}$ for a weak SPDC pump beam at another wavelength~\cite{Rarity2005,Ramelow2011,Jennewein2014,Jennewein2015}.   Presuming that the induced $\chi^{(2)}$ value enables unity-efficiency conversion of the $|0,0,2\rangle$ input state to a $|2,2,0\rangle$ output state, a $K$-level cascade of these UPDC systems then enables unity-efficiency preparation of the $\ket{2^K,2^K}$ dual-Fock polarization state from the $|0,0,2\rangle$ input state, as shown in Sec.~IV of~\cite{groverAppendix}. This method requires efficient preparation of the two-photon Fock-state pump, which is experimentally challenging at present.  Theoretical suggestions for such Fock-state preparation include Refs.~\cite{Motes2016,Santos2001}.  Microwave generation experiments include Refs.~\cite{Hofheinz2008, Premaratne2017}, which could yield two-photon optical pumps by means of microwave-to-optical quantum-state frequency conversion (QSFC).  See Refs.~\cite{Kumar1990,Marius2006,Marius2004,Zaske2012,Zeilinger2012} for optical-to-optical QSFC.

In conclusion, we have studied the quantum theory of SPDC with single-mode signal, idler, and pump beams and Fock-state pumps. We found that the efficiency of converting pump photons into signal-idler photon pairs is unity only for the single-photon pump.  In order to transcend this fundamental limit, we proposed using amplitude amplification, analogous to Grover search, of the completely-converted state by interlacing SPDC processes with NSGs.  Our method can realize unity-efficiency conversion,  with nonlinear crystals of the appropriate lengths, for all pump-photon numbers, but the required crystal-length precision becomes increasingly demanding with increasing pump-photon number.  Nevertheless, unity-efficiency conversion should be possible for pump-photon numbers up to 5, even if the same crystal length is used for all Grover iterations.  

M.Y.N., F.N.C.W.,  and J.H.S. acknowledge support from Office of Naval Research grant number N00014-13-1-0774 and Air Force Office of Scientific Research grant number FA9550-14-1-0052. B.C.S. acknowledges funding provided by Natural Sciences and Engineering Research Council of Canada, Alberta Innovations, and by the Institute for Quantum Information and Matter, an National Science Foundation Physics Frontiers Center (NSF Grant PHY-1125565) with support of the Gordon and Betty Moore Foundation (GBMF-2644). M.Y.N. and B.C.S. also thank the Aspen Center for Physics for hosting the Quantum Algorithm Winter School.

\begin{widetext}

\begin{center}
{\Large\bf Supplemental  Material} 
\end{center}

\section{Quantum Theory of Parametric Down-Conversion}

In this section we present analytic results for the quantum dynamics of parametric down-conversion with single-mode signal, idler, and pump beams whose joint state is evolving in the $n$-pump-photon subspace where $n\le 4$.  We start from 
the three-wave-mixing interaction Hamiltonian 
\begin{equation}
	\hat{H}
		=\text{i}\hbar \kappa\!\left(\hat{a}_s^{\dagger}\hat{a}_i^{\dagger}\hat{a}_p - \hat{a}_p^{\dagger}\hat{a}_s\hat{a}_i\right),
\label{HSPDC}
\end{equation}
where $\kappa$ is assumed to be real valued, and seek solutions to the Schr\"{o}dinger equation
\begin{equation}
\text{i}\hbar\ket{\dot{\Psi}(t)}=\hat{H} \ket{\Psi(t)}, \mbox{ for $t\ge 0$,}
\end{equation}
subject to the initial condition $|\Psi(0)\rangle$ on the joint state of the signal, idler, and pump.  In general, this initial condition can be decomposed into components that lie within subspaces spanned by $\{|0,0,n\rangle, |1,1,n-1\rangle,\ldots,|n,n,0\rangle\}$, where $|n_s,n_i,n_p\rangle$ denotes a state containing $n_s$ signal photons, $n_i$ idler photons, and $n_p$ pump photons, i.e.,
\begin{equation}
|\Psi(0)\rangle = \sum_{n=0}^\infty c_n|\Psi_n(0)\rangle, \quad\mbox{ where }\sum_{n=0}^\infty |c_n|^2 = 1,
\end{equation}
and 
\begin{equation}
|\Psi_n(0)\rangle = \sum_{k=0}^n f^{(n)}_k(0) |k,k,n-k\rangle, \quad\mbox{ with }\sum_{k=0}^n|f^{(n)}_k(0)|^2 = 1.
\end{equation}  

Schr\"{o}dinger evolution occurs independently within each of these $n$-pump-photon subspaces according to the following coupled ordinary differential equations:  
\begin{align}
	\dot{f}_k^{(n)}(t)=
		\begin{cases}
			-\kappa\sqrt{n}f_{k+1}^{(n)}(t), & k=0\\[.05in]
			\kappa\left[k\sqrt{n-k+1}f_{k-1}^{(n)}(t)- (k+1)\sqrt{n-k}f_{k+1}^{(n)}(t)\right], &  k=1,2,\ldots, n-1\\[.05in]
			\kappa nf_{n-1}^{(n)}(t),  &k=n.
		\end{cases}
\label{ode}
\end{align}
Equations~\eqref{ode} have closed-form solutions for $n\leq 4$, given $\{f_k^{(n)}(0) : 0\le k \le n\}$, but the Abel-Ruffini theorem tells us that no such analytic solutions are possible for $n \ge 5$.  In the remainder of this section we explore the implications of the closed-form solutions with respect to the efficiency of converting pump photons to signal-idler photon pairs.   We assume that the $\{f_k^{(n)}(0)\}$ are real valued, so that the $\{f_k^{(n)}(t)\}$ are also real valued.  (Inasmuch as our principal interest is in the initial condition $f_k^{(n)}(0) = \delta_{k0}$, where $\delta_{k0}$ is the Kronecker delta, there is little loss of generality in making the real-valued assumption.)

For the one-pump-photon subspace, Eqs.~\eqref{ode} imply that  \begin{align}\label{single1}
	&f_0^{(1)}(t) = f_0^{(1)}(0)\cos(\kappa t),\\[.05in]\label{singlephotonsolution}
	&f_1^{(1)}(t) = f_1^{(1)}(0)\sin(\kappa t).
\end{align}
It follows that single-mode down-conversion with a one-photon pump can achieve unity-efficiency conversion to a signal-idler photon pair, i.e., the initial state $|0,0,1\rangle$ is completely converted to $|1,1,0\rangle$ when $t= \pi/2\kappa$.  We shall see below that 
such unity-efficiency conversion is \emph{not} possible with SPDC in the $n$-pump-photon subspaces for $n=2,3,4$.    

For the two-pump-photon subspace, Eqs.~(\ref{ode}) yield the general solution
\begin{align} \nonumber
f_0^{(2)}(t)=& \frac{2}{3} \sqrt{\frac{3}{1+2m^2}} \left[m+ \frac{1}{2}\cos\!\left(\kappa \sqrt{6}\, t+\phi_0\right) \right],\\[.05in]\label{solution2Photons}
f_1^{(2)}(t)= & \frac{1}{\sqrt{1+2m^2}} \sin\!\left(\kappa \sqrt{6}\, t+\phi_0\right),\\[.05in]
f_2^{(2)}(t)= & \frac{\sqrt{6}}{3\sqrt{1+2m^2}} \left[m-\cos\!\left(\kappa \sqrt{6}\, t+\phi_0\right) \right],\nonumber
\end{align}
where $m$ and $\phi_0$ are determined by the initial conditions $\{f_k^{(2)}(0) : 0\le k \le 2\}$.  

Figure~\ref{3Df2} shows five $\{f_k^{(n)}(t)\}$ trajectories that were obtained from Eqs.~(\ref{solution2Photons}) using the initial conditions $m$ = 0, 0.4, 1, 2, and 4, all with $\phi_0=0$.   The state evolution for $f_k^{n)}(0) = \delta_{k0}$, given by Eqs.~(\ref{solution2Photons}) with $m=1$ and $\phi_0 = 0$, leads to a maximum conversion efficiency
\begin{align}
\mu_2=\max_t  \sum_{k=1}^2 \frac{k|f_k^{(2)}(t)|^2}{2}\approx 0.89,
\end{align}
with virtually all of the conversion being to the $|2,2,0\rangle$ state, because $|f_2^{(2)}(t_{\rm opt})|^2 \gg |f_1^{(2)}(t_{\rm opt})|^2$, where $t_{\rm opt}$ is the interaction time that maximizes $\mu_2$.

\begin{figure}
\begin{center}
\includegraphics[width=0.35\linewidth]{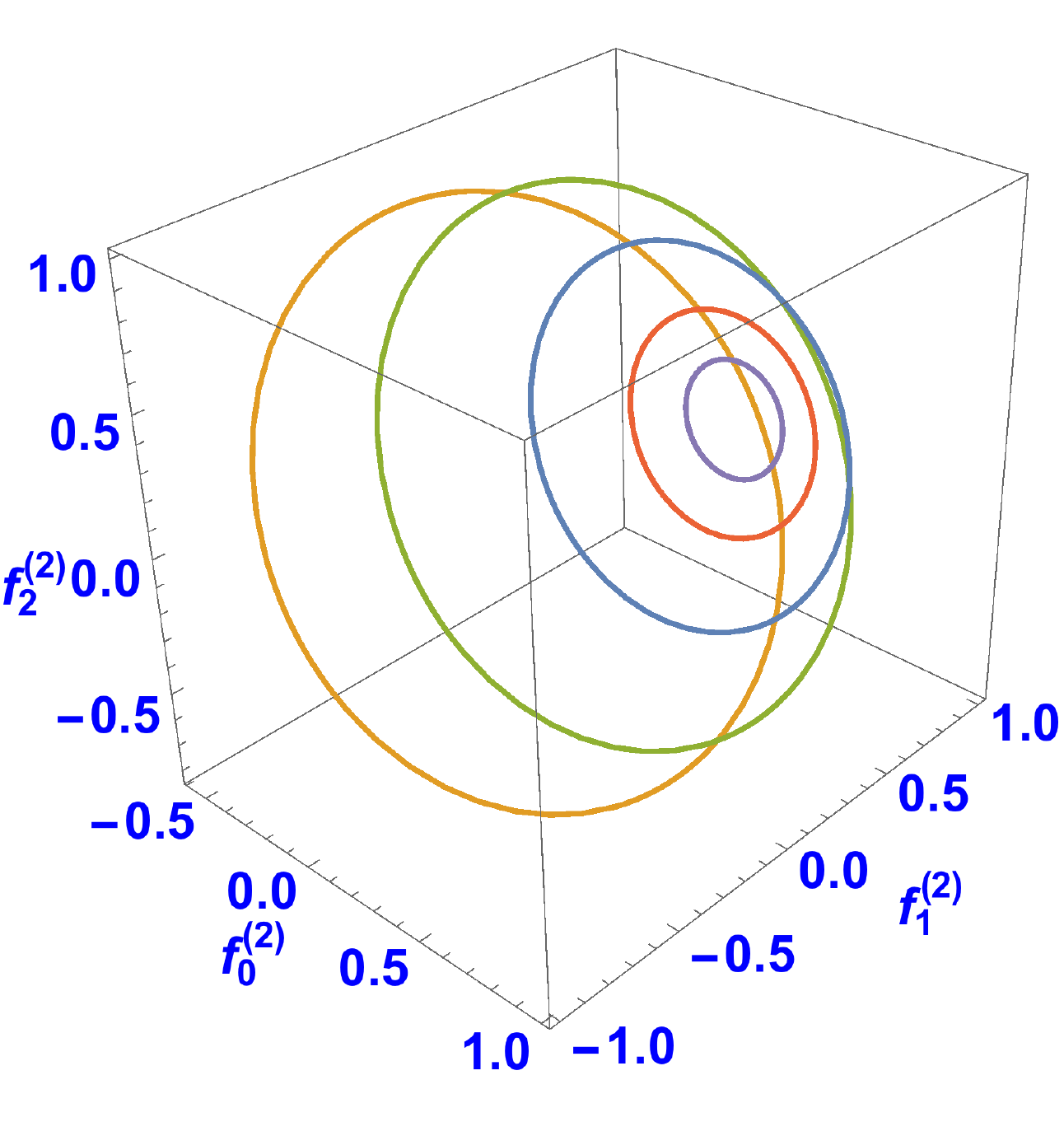}
\caption{Trajectories of $\{f_k^{(2)}(t)\}$ obtained from Eqs.~\eqref{solution2Photons} for $\phi_0 = 0$ with $m=0$ (orange), 0.4 (green),1 (blue), 2 (red), and 4 (purple). These trajectories form circles on the unity-radius sphere that is centered at the origin, $(f_0^{(2)},f_1^{(2)},f_2^{(2)}) = (0, 0, 0)$.   The circles lie in planes that are perpendicular to the line $f_1^{(2)} = \sqrt{2}\,f_2^{(2)}$ in the $f_0^{(2)} = 0$ plane. \label{3Df2}}
\end{center}
\end{figure}

Equations~\eqref{ode}'s solutions for the three-pump-photon subspace take the general form
\begin{align}\nonumber
	f_0^{(3)}(t) &= \left[B_+\cos(\omega_+ \kappa t +\phi_1)+ B_-\cos(\omega_- \kappa t+\phi_2)\right],\\[.05in]\nonumber
f_1^{(3)}(t) &= \frac{1}{\sqrt{3} }\left[B_+\omega_+\sin(\omega_+ \kappa t+\phi_1)+ B_-\omega_-\sin(\omega_- \kappa t+\phi_2)\right],\\[.05in]\label{3photons}
f_2^{(3)}(t)&=\sqrt{\frac{6}{73}}\,\left[\cos(\omega_+ \kappa t+\phi_1)-\cos(\omega_- \kappa t+\phi_2)\right],\\[.05in]\nonumber
f_3^{(3)}(t)&= 6\sqrt{\frac{3}{146}}\left[\frac{\sin(\omega_+ \kappa t+\phi_1)}{\omega_+}- \frac{\sin(\omega_- \kappa t+\phi_2)}{\omega_-}\right],\nonumber
\end{align}
with $\omega_\pm=\sqrt{10\mp\sqrt{73}}$ and the remaining constants being determined by the initial conditions $\{f_k^{(3)}(0)\}$.  Here, the irrationality of $\omega_+/\omega_-$ implies that the $\{f_k^{(3)}(t)\}$ evolve in an aperiodic manner.  We illustrate this aperiodic behavior in Fig.~\ref{3pohton3d}, where we have plotted $f_0^{(3)}(t)$, $f_3^{(3)}(t)$, and $f_\perp^{(3)} \equiv \sqrt{1-|f_0^{(3)}(t)|^2 - |f_3^{(3)}(t)|^2}$ for $0 \le t \le 30\pi/\kappa\omega_+$ and initial condition $f_0^{(3)}(0) = 1$, which corresponds to $\phi_1=\phi_2 =0$ and $B_{\pm} = (\sqrt{73}\pm 7)/2\sqrt{73}$.  In this case we find that the maximum conversion efficiency to the $|3,3,0\rangle$ completely-converted state is $\max_t |f_3^{(3)}(t)|^2 \approx 0.40,$ while the maximum conversion efficiency is 
\begin{align}
\mu_3=\max_t  \sum_{k=1}^3 \frac{k|f_k^{(3)}(t)|^2}{3}\approx 0.89.
\end{align}

\begin{figure}
\includegraphics[width=0.4\linewidth]{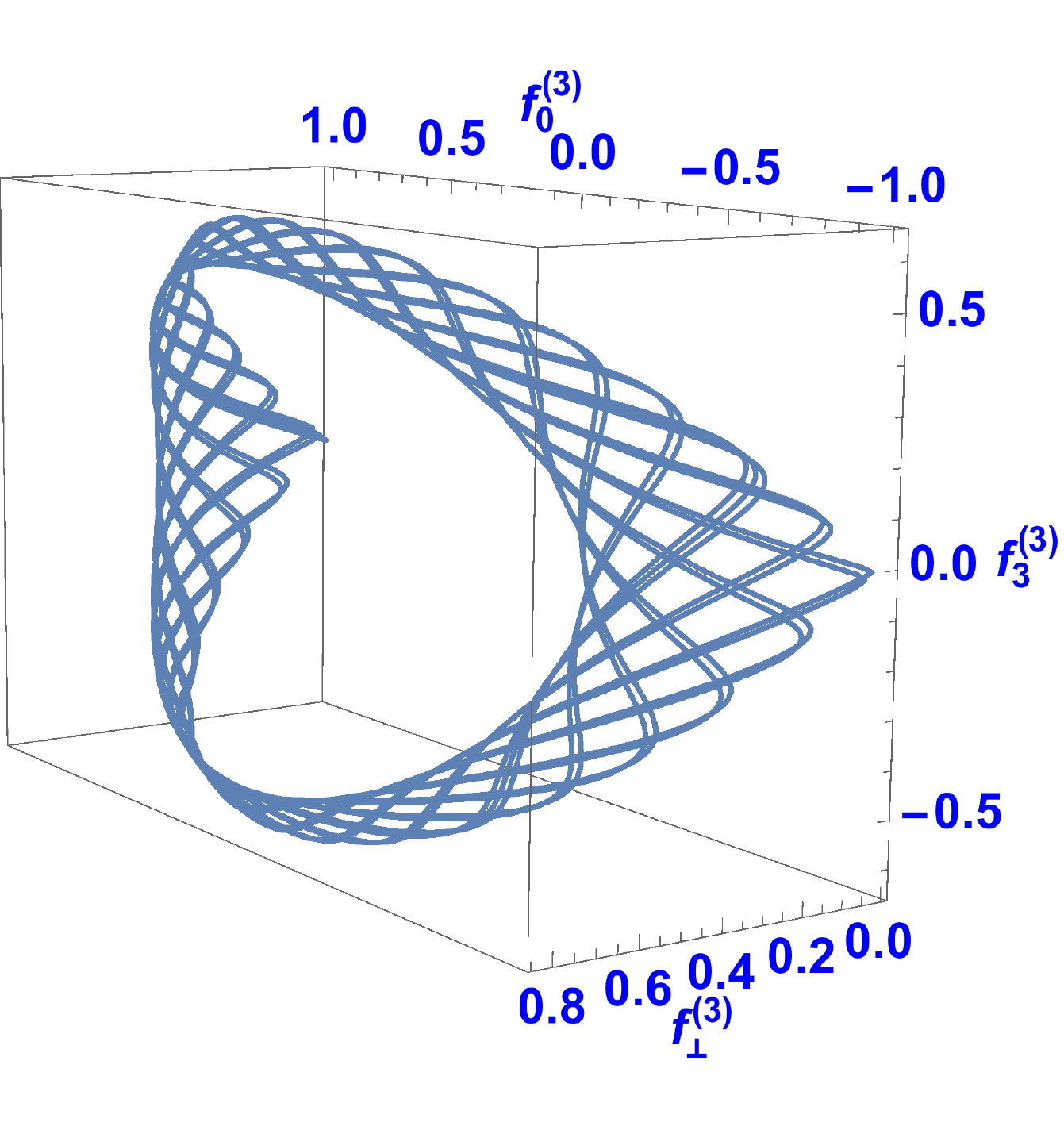}
\caption{Trajectories of $\left\{f_0^{(3)}(t),f_3^{(3)}(t),f_\perp^{(3)} \equiv \sqrt{1-|f_0^{(3)}(t)|^2 - |f_3^{(3)}(t)|^2}\right\}$ for $0 \le t \le 30\pi/\kappa\omega_+$ and initial condition $f_0^{(3)}(0) = 1$.} \label{3pohton3d}
\end{figure}

For the four-pump-photon subspace, we limit our attention to the behavior of $f_0^{(4)}(t)$ and $f_4^{(4)}(t)$ when the initial condition is $f_k^{(4)}(0) = \delta_{k0}$:  
\begin{align}\nonumber
& f_0^{(4)}(t)
		= \frac{C}{17/41 + m^2/1168992}
			\left[B_+\cos(\omega_+ \kappa t)
				+B_-\cos(\omega_- \kappa t) + m\right],\\ \label{xyz2}
& f_4^{(4)}(t)
		= -\frac{6\sqrt{6}C}{17/41+m^2/1168992} \left[\omega_-^2\cos(\omega_+ \kappa t)- \omega_+^2\cos(\omega_- \kappa t) - m/24\right],
\end{align} 
where $\omega_\pm= \sqrt{25\mp 3\sqrt{33}}, m=144\sqrt{33}, C=1/246\sqrt{33},$ and  $B_\pm=51\sqrt{33}\pm261$.  Here, the maximum conversion efficiency to the full-converted $\ket{4,4,0}$ state is $\max_t |f_4^{(4)}(t)|^2 \approx 0.74$, which is lower than the two-pump-photon subspace's maximum conversion efficiency to its fully-converted state but higher than that for the three-pump-photon subspace.  On the other hand, the maximum conversion efficiency of the four-photon pump input is $\mu_4\approx 0.86,$ which is lower than that for both the two-pump-photon and three-pump-photon subspaces.

\section{Unity-Efficiency Conversion in the Limit of High Pump-Photon Numbers}\label{LargePhoton}
In this section we provide a proof by induction that amplitude amplification can achieve an arbitrarily-close-to-unity efficiency for converting the input state $|0,0,n\rangle$ to the completely-converted output state $|n,n,0\rangle$ in the large-$n$ limit.  We preface our induction proof by justifying the assertion that passing the input state through a length-$L_0$, type-II phase-matched $\chi^{(2)}$ crystal (equivalent to an interaction time $t_0 = L_0/v$) yields the UPDC procedure's initial state
\begin{equation}
|\Psi_0\rangle = \cos(\theta_g/2)|0\rangle + \sin(\theta_g/2)|1\rangle,
\label{initialstate}
\end{equation}
for $0 < \theta_g \ll 1$, where $|1\rangle \equiv |n,n,0\rangle$ and $|0\rangle$ is a normalized state satisfying $\langle 1|0\rangle = 0$.  For $\kappa \delta t \ll 1$, Eqs.~(\ref{ode}) yield
\begin{align}
f_1^{(n,0)}(\delta t)&=\kappa\sqrt{n}\int_0^{\delta t}f_0^{(n,0)}(t)\,{\rm d}t=\sqrt{n}\,\kappa \delta t,\\
f_2^{(n,0)}(\delta t)&= 2\kappa\sqrt{n-1}\int_0^{\delta t}f_1^{(n,0)}(t)\,{\rm d}t= \sqrt{n(n-1)}\,(\kappa\delta t)^2,\\
&\vdots\\ \label{fn-1}
f_{n-1}^{(n,0)}(\delta t)&=\sqrt{n!}\, (\kappa\delta t)^{n-1},\\
f_{n}^{(n,0)}(\delta t)&=\sqrt{n!}\,(\kappa \delta t)^n,
\end{align}
to lowest order in $\delta t$.  Setting $t_0 = \delta t$ then shows that we can realize Eq.~(\ref{initialstate}) with $\sin(\theta_g/2) = \sqrt{n!}\,(\kappa t_0)^n$.  

At this point we begin the induction proof in earnest.  We must first show that, after applying the $U_{\rm NSG}^{(n)}$ gate to $|\Psi_0\rangle$ to obtain the state $|\Psi_1\rangle$, there is an SPDC crystal length $L_1$ (equivalent to an interaction time $t_1 = L_1/v$) which will produce
\begin{equation}
|\Psi_1'\rangle = \cos(3\theta_g/2)|0\rangle + \sin(3\theta_g/2)|1\rangle.
\label{firstGroverState}
\end{equation}
as the first Grover iteration's output state, with $|0\rangle$ being a normalized state satisfying $\langle 1|0\rangle = 0$.
Consider an interaction time $t_1'$ satisfying $\kappa t_1' \ll 1$.  We have 
\begin{equation}
 f_{n-1}^{(n,0)} (t_{0})=\sin(\theta_g/2)^{(n-1)/n} (n!)^{1/2n} 
\approx\sin(\theta_g/2)\sqrt{n/e},
\end{equation}
where we have used the Stirling approximation for $n!$ and $\sin(\theta_g/2)^{(n-1)/n} \approx \sin(\theta_g/2)$ for $n\gg 1$.  Using this result we find that
\begin{align}
f_n^{(n,1)} (t_1')&=-f_n^{(n,0)} (t_{0}) + \int_{t_0}^{t_0 + t_1'} \kappa n f_{n-1}^{(n,0)} (t)\, {\rm d}t\\ \label{E27}
&\approx -\sin(\theta_g/2) + \int_{t_0}^{t_0 + t_1'}\kappa \sin(\theta_g/2)\sqrt{n/e}\, {\rm d}t\\
&=\sin(\theta_g/2)(\sqrt{n/e}\,\kappa t_1'-1).
\end{align}
Because $0< \sin(3\theta_g/2) < 3\sin(\theta_g/2)$ for $0<\theta_g \ll 1$, it follows that having $\kappa t_1'\ll 1$ and $\kappa t_1' \geq 4\sqrt{e/n}$ ensures there is a $t_1 < t_1'$ such that Eq.~(\ref{firstGroverState}) holds.  

Next, we assume that 
\begin{equation}
|\Psi_m'\rangle = \cos[(2m+1)\theta_g/2]|0\rangle + \sin[(2m+1)\theta_g/2]|1\rangle,
\label{mthGroverState}
\end{equation}
for $m>1$,  is the $m$th Grover iteration's output state, where $|0\rangle$ is a normalized state satisfying $\langle 1|0\rangle = 0$.  Our induction proof will be complete if we can show that 
\begin{equation}
|\Psi_{m+1}'\rangle = \cos[(2m+3)\theta_g/2]|0\rangle + \sin[(2m+3)\theta_g/2]|1\rangle,
\label{(m+1)thGroverState}
\end{equation}
with $|0\rangle$ being a normalized state satisfying $\langle 1|0\rangle = 0$, is the $(m+1)$th Grover iteration's output state.

Using $f^{(n,m+1)}_k(0) = (-1)^{\delta_{kn}}f^{(n,m)}_k(t_m)$, which holds for $m>1$, Eqs.~(\ref{ode}) give us 
\begin{align}
f_{n-1}^{(n,m+1)}( \delta t) &= \kappa[\sqrt{2}\,(n-1)f_{n-2}^{(n,m)}(t_m) + nf_n^{(n,m)}(t_m)]\kappa\delta t + f_{n-1}^{(n,m)}(t_m)\\\label{equivalenceNN+1}
&= f_{n-1}^{(n,m)}(t_m +\delta t)  + 2nf_n^{(n,m)}(t_m)\kappa \delta t,
\end{align}
and
\begin{equation}
f_n^{(n,m+1)}(\delta t)= -f_n^{(n,m)}(t_m ) + \int_{0}^{\delta t} \kappa n f_{n-1}^{(n,m+1)}(t)\,{\rm d}t,
\label{fnm+1}
\end{equation}
for $\kappa \delta t \ll 1$.
Another use of Eqs.~(\ref{ode}) with $\kappa\delta t \ll 1$ plus Eq.~(\ref{equivalenceNN+1}) then leads to 
\begin{align}
\int_{0}^{ \delta t} \kappa n f_{n-1}^{(n,m+1)}(t)\,{\rm d}t&=\int_{0}^{ \delta t} \kappa nf_{n-1}^{(n,m)}(t_m + t)\,{\rm  d}t +\int_{0}^{ \delta t} 2\kappa^2 n^2f_n^{(n, m)}(t_m) t\,{\rm  d}t\\
&= f_n^{(n,m)}(t_m +\delta t) -f_n^{(n,m)}(t_m )+(n\kappa \delta t)^2f_n^{(n,m)}(t_m).
\end{align}
Substituting this result into Eq.~(\ref{fnm+1}), we have that
\begin{align}
f_n^{(n,m+1)}(\delta t) &=-2f_n^{(n,m)}(t_m )+f_n^{(n,m)}(t_m +\delta t) + (n\kappa \delta t)^2f_n^{(n,m)}(t_m)\\
&\geq f_n^{(n,m)}(t_m )[(n \kappa\delta t)^2 -2],
\end{align}
where $f^{(n,m)}_n(t_m) = \sin[(2m+1)\theta_g/2] >0$, and the continuity of the Schr\"{o}dinger evolution plus $\kappa \delta t \ll 1$ ensures that $f_n^{(n,m)}(t_m +\delta t) >0$.  Now we see that 
\begin{equation}
f_n^{(n,m+1)}(\delta t) \geq  \sin[(2m+3)\theta_g/2]
\end{equation}
prevails if
\begin{equation}
\kappa \delta t \geq \frac{\sqrt{2+\frac{\displaystyle \sin[(2m+3)\theta_g/2]}{\displaystyle\sin[(2m+1)\theta_g/2]}}}{n},
\end{equation}
and this can be satisfied with $\kappa\delta t \ll 1$ if 
\begin{equation}
n \gg \sqrt{2+\frac{\displaystyle \sin[(2m+3)\theta_g/2]}{\displaystyle\sin[(2m+1)\theta_g/2]}}\,.
\end{equation}

Because $0< \sin(3\theta_g/2) < 3\sin(\theta_g/2)$ for $0<\theta_g \ll 1$, and $\sin[(2m+1)\theta_g/2]$ is monotonically decreasing with increasing $m$, the preceding condition on $n$ is met if $n \gg \sqrt{5}$.  So, choosing $n$ large enough we can find a $t_{m+1}$ that provides the amplitude amplification needed to complete the induction proof.  Thus, with $M$ being the largest integer satisfying $(2M+1)\theta_g <\pi$, we can get a $\sin^2[(2M+1)\theta_g]$ conversion efficiency, from the input state $|0,0,n\rangle$ to the fully-converted state $|n,n,0\rangle$, and this conversion efficiency can be made arbitrarily close to unity for small enough $\theta_g$.  Furthermore, choosing $\theta_g \simeq 1/\sqrt{n}$, for $n\gg 1$, we have that $M$ is $O(\sqrt{n})$, as expected for Grover search. 

\section{Grover-search example:  Two-pump-photon subspace}\label{groverExample}

\begin{figure} [thb]
\begin{center}
\includegraphics[width=0.4\linewidth]{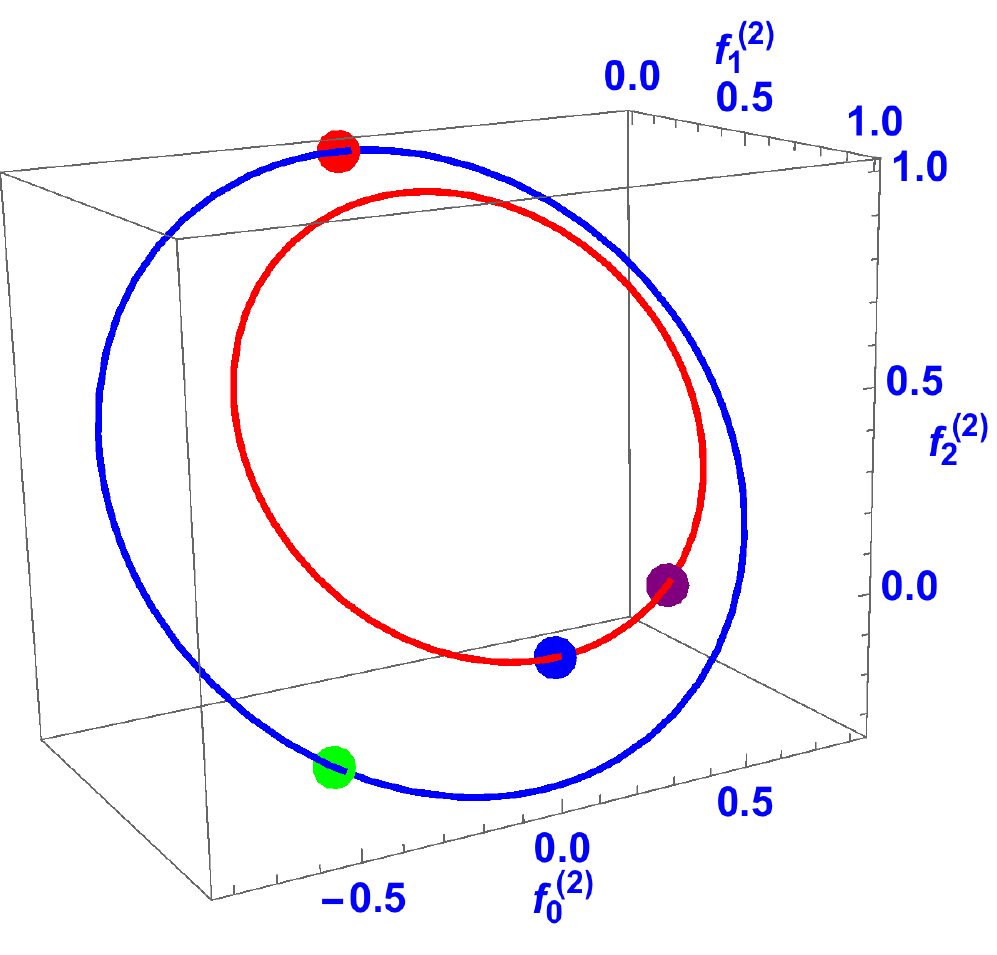}
 \caption{3D plot showing the UPDC procedure in the two-pump-photon subspace that realizes unity-efficiency conversion from the $|0,0,2\rangle$ input state, shown as the blue dot, to the $|2,2,0\rangle$ final state, shown as the red dot, in a single Grover iteration.   The UPDC procedure's initial state $|\Psi_0\rangle$, prepared by passing the input state through a type-II phase-matched $\chi^{(2)}$ crystal for an interaction time $t_0 = 0.976/\kappa\sqrt{6}$, is shown by the purple dot that is obtained by evolution around the red circle from the blue dot.   Sign flip on the marked state ($|0,0,2\rangle$) transforms the $|\Psi_0\rangle$ state to $|\Psi_1\rangle$, which is indicated by the green dot. Rotation toward the marked state by passing $|\Psi_1\rangle$  through a type-II phase-matched $\chi^{(2)}$ crystal for an interaction time $t_1 = (\pi-0.626)/\kappa\sqrt{6}$ leads to evolution around the blue circle to $|\Psi_1'\rangle$ indicated by the red dot, which is the desired output state $|2,2,0\rangle$.  \label{2DgroverF2}  }
\end{center}
\end{figure}
Here we supplement the large-$n$ proof from Sec.~\ref{LargePhoton} by presenting an explicit demonstration of complete conversion to the fully-converted state for the two-pump-photon subspace.  In particular, using Eqs.~(\ref{solution2Photons}), we show that the four-step UPDC procedure described in the main paper realizes complete conversion in a single Grover iteration.
\begin{enumerate}
\item[I.]\label{step1} \textit{Initialization: }The input state $|0,0,2\rangle$, shown as the blue dot in Fig.~\ref{2DgroverF2}, undergoes a duration-$t_0$ interaction in the type-II phase-matched $\chi^{(2)}$ crystal to yield, via Eqs.~(\ref{solution2Photons}) with $m=1$ and $\phi_0 = 0$, the UPDC procedure's initial state
\begin{equation}
|\Psi_0\rangle = \frac{2}{3}\left[1+\frac{1}{2}\cos\!\left(\kappa\sqrt{6}\,t_0\right)\right] \ket{0,0,2}+ \frac{1}{\sqrt{3}}\sin\!\left(\kappa\sqrt{6}\,t_0\right)\ket{1,1,1} + \frac{\sqrt{2}}{3}\left[1-\cos\left(\kappa\sqrt{6}\,t_0\right)\right] \ket{2,2,0}.\label{originalF2Grover}
\end{equation}
In order to achieve complete conversion in a single Grover iteration, we choose $t_0 = 0.976/\kappa\sqrt{6}$, which leads to $|\Psi_0\rangle$ being the purple dot in Fig.~\ref{2DgroverF2} obtained from duration-$t_0$ evolution around the red circle from the blue dot in that figure.

\item[II.]\textit{Sign flip on the marked state: }Applying the $U_{\rm NSG}^{(2)}$ gate to the $|\Psi_0\rangle$ obtained with $t_0 = 0.976/\kappa\sqrt{6}$ yields
\begin{equation}
|\Psi_1\rangle = \frac{2}{3}\left[1+\frac{1}{2}\cos(0.976)\right] \ket{0,0,2}+ \frac{1}{\sqrt{3}}\sin(0.976)\ket{1,1,1} - \frac{\sqrt{2}}{3}[1-\cos(0.976)] \ket{2,2,0},
\end{equation}
which corresponds to transitioning from the purple dot on the red circle to the green dot on the blue circle in Fig.~\ref{2DgroverF2}.

\item[III.]\textit{Rotation toward the marked state: }Using the $|\Psi_0'\rangle$ obtained with $t_0 = 0.976/\kappa\sqrt{6}$ as the input to a duration-$t_1$ interaction in a $\chi^{(2)}$ crystal implies that the initial conditions Eqs.~(\ref{solution2Photons}) use for that evolution are $m=1/2$ and   
$\phi_0 = 0.626$.  With those initial conditions Eqs.~(\ref{solution2Photons}) now give us
\begin{equation}
\ket{\Psi_1^\prime}= \frac{\sqrt{2}}{3}[1+\cos(\kappa\sqrt{6}\,t_1+\phi_0)] \ket{0,0,2} + \frac{\sqrt{2}}{\sqrt{3}}\sin(\kappa\sqrt{6}\,t_1+\phi_0)\ket{1,1,1}   + \frac{1}{3}[1-2\cos(\kappa\sqrt{6}\, t_1+\phi_0)]\ket{2,2,0}.
\end{equation}
To obtain complete conversion we choose $t_1 = (\pi-\phi_0)/\kappa\sqrt{6}$, which reduces $|\Psi_1'\rangle$ to $|2,2,0\rangle$, as shown by the red dot in Fig.~\ref{2DgroverF2} obtained from duration-$t_1$ evolution around the blue circle in that figure.

\item[IV.]\label{step4}\textit{Termination: }Complete conversion having been achieved, the UPDC procedure's Grover iterations terminate after a single iteration.

\end{enumerate}

Figure~\ref{2photonGroverScheme} is a schematic for realizing the two-pump-photon UPDC procedure's Steps I through IV using the nondeterministic NSG proposed in Ref.~\cite{Knill2001}. The corresponding schematic for the deterministic-NSG version of two-photon-pump UPDC is the Level~1 unit cell in Fig.~\ref{cascade}.

\begin{figure} [thb]
\begin{center}
\includegraphics[width=0.5\linewidth]{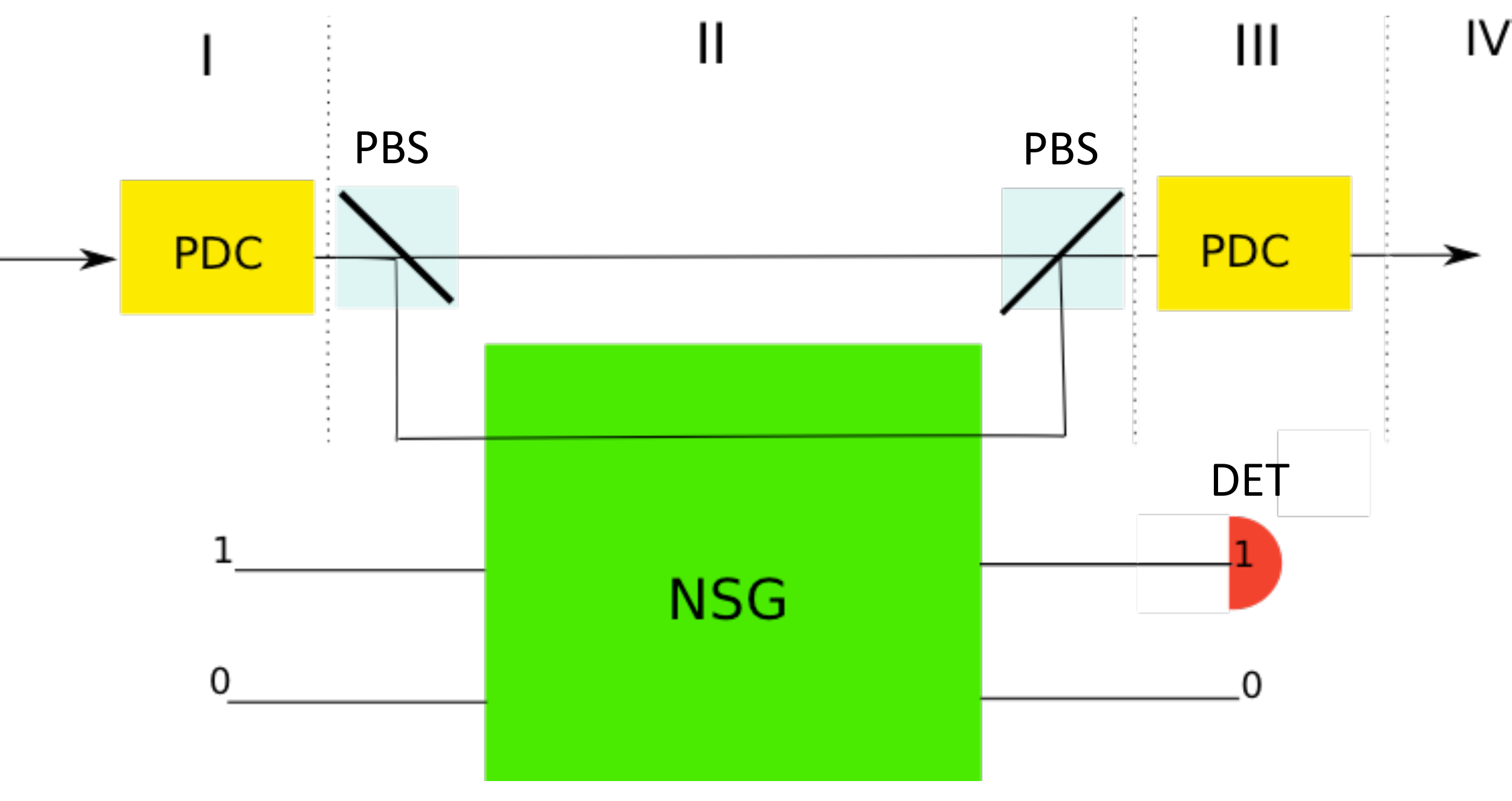}
\caption{Schematic for the two-pump-photon UPDC procedure using a nondeterministic NSG. Dotted lines separate the procedure's Steps I through IV, whose descriptions were given earlier in Sect. \ref{step1}.  The PDC blocks are parametric down-converters and the NSG block is the nondeterministic nonlinear sign gate from Ref.~\cite{Knill2001}.   The PBS blocks are polarization beam splitters.  One directs the signal photons emerging from the first PDC into the NSG, and the other recombines the signal photons emerging from the NSG with idler and pump photons at the input to the second PDC. The upper and lower ancilla rails entering the NSG are prepared in the single-photon Fock state and the vacuum state---here denoted $1$ and $0$---and an array of beam splitters within the NSG block (omitted here for simplicity) performs the unitary transformations described in Ref.~\cite{Knill2001} for nondeterministic NSG realization.   Thus, when the first PDC's input is in the $\ket{0,0,2}$ state and the detector (DET) counts one photon, then the output signal-idler-pump joint state will be $ \ket{2,2,0}.$ \label{2photonGroverScheme}  }
\end{center}
\end{figure}
\section{Dual-Fock state Generation via Cascaded Two-Pump-Photon UPDC }

An interferometer whose two input ports are illuminated by the dual-Fock state $\ket{n,n}$ enjoys a quadratic improvement in phase-sensing precision over a coherent-state system of the same average photon number, thus achieving Heisenberg-limited performance~\cite{Holland1993}.   The signal and idler outputs from SPDC, however, are in a thermal distribution of~$\ket{n,n}$ states that eradicates this advantage~\cite{Sanders1995}.  We show in this section that cascaded two-pump-photon UPDC can produce a particular class of large-$n$ dual-Fock states.  In Sec.~\ref{LargePhoton} we proved that large-$n$ dual-Fock states can be generated, in principle, via $n$-pump-photon UPDC, but that approach requires $U_{\rm NSG}^{(n)}$ gates for which there is no known deterministic realization, and their nondeterministic realization has $O(1/n)$ success-probability scaling.  More generally, the state-of-the-art proposal for preparing a large-$n$ dual-Fock state is nondeterministic~\cite{Motes2016}.   Generating a particular class of large-$n$ dual-Fock states via cascaded two-pump-photon UPDC, on the other hand, is a deterministic procedure if its UPDC elements employ $U_{\rm NSG}^{(2)}$ gates realized with nonlinear optics.   
\begin{figure}[h]
\includegraphics[width=0.8\linewidth]{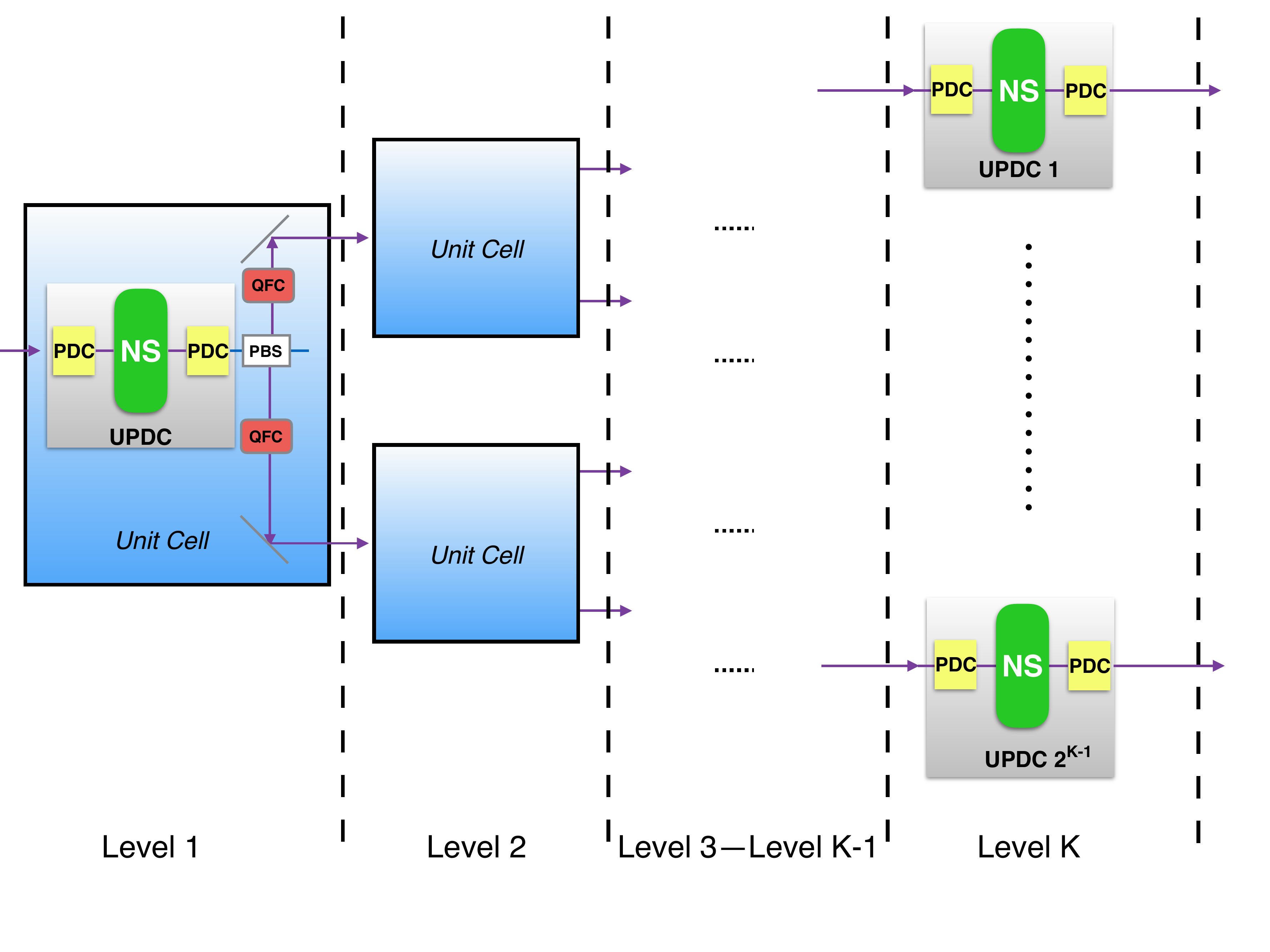}
\caption{Schematic setup for generating dual-Fock states via $K$-level cascaded UPDC.  Level $k$, for $1\le k \le K-1$, employs $2^{k -1}$ unit cells, each consisting of a UPDC unit followed by a polarization beam splitter (PBS) and two quantum-state frequency converters (QFCs), while in Level $K$ only the UPDC units are employed.  Each UPDC unit contains two parametric down-conversion~(PDC) crystals and a deterministic $U^{(2)}_{\rm NSG}$ nonlinear sign gate (NSG) that give 100\% efficient conversion of a $|0,0,2\rangle$ input state to the $|2,2,0\rangle$ output state. For Levels $1\le k \le K-1$, the PBS first separates the signal and idler into distinct spatial modes, whose photons are then converted, by QFCs, into the two-pump-photon input states needed for the next level in the cascade.  The $2^K$ signal and $2^K$ idler photons at the output of Level $K$ can be combined, using a delay-and-switch procedure (not shown), into a single spatial mode, as described in the text.}
\label{cascade} 
\end{figure}

Figure~\ref{cascade} shows a $K$-level version of our cascaded two-pump-photon SPDC scheme for generating dual-Fock states.  Its fundamental building block is a unit cell comprised of a $t_0 = 0.976/\kappa\sqrt{6}$ interaction time, type-II phase-matched $\chi^{(2)}$ crystal, a deterministic $U_{\rm NSG}^{(2)}$ gate, a $t_1 = (\pi-0.626)/\kappa\sqrt{6}$ interaction time, type-II phase-matched $\chi^{(2)}$ crystal, a polarization beam splitter, and two quantum-state frequency converters.  From Sec.~\ref{groverExample} we know that sandwiching the $U_{\rm NSG}^{(2)}$ gate between a unit cell's two down-conversion crystals will take a two-photon pump at frequency $\omega_p$ and convert it to two pairs of orthogonally-polarized signal and idler photons at frequencies $\omega_s$ and $\omega_i$, respectively.  The signal and idler photons are separated into distinct spatial modes by the polarization beam splitter, after which they individually enter quantum-state frequency converters~\cite{Kumar1990, Marius2004,Marius2006,Zaske2012,Zeilinger2012}.  The frequency converters perform 100\%-efficiency conversion of their two-photon inputs to two-photon outputs at the pump frequency and in the polarization needed for pumping the next cascade level's down-conversion crystals.  The final level in a $K$-level cascade, however, does not use polarization beam splitters or quantum-state frequency converters. Its outputs are $2^{K-1}$ spatial modes each containing a $|2,2,0\rangle$ signal-idler-pump state, making $|2,2,0\rangle^{\otimes 2^{K-1}}$ the joint state of these spatial modes.  

The preceding $2^{K-1}$ signal-idler outputs from the $K$th cascade level can now be combined into a single spatial mode by the following delay-and-switch procedure.  Suppose that these outputs are all in a common temporal mode, $\psi(t)$, that is time limited to $|t| \le T/2$.  For $1 \le \ell \le 2^{K-1}$, we delay the $\ell$th spatial mode by $\ell\Delta T$, where $\Delta T >T$.  We then use an optical switch yard to coherently combine the $2^{K-1}$ delayed signal-idler beams into a single spatial mode containing $2^K$ signal photons and $2^K$ idler photons.   For applications in which only polarization---not temporal mode---matters, the single spatial-mode we have created with our delay-and-switch procedure will be in the $|2^K,2^K\rangle$ state, where the first and second entries denote the signal-frequency, signal-polarization photon number and idler-frequency, idler-polarization photon number, respectively.

\end{widetext}


\begin{thebibliography}{99}
\bibitem{footnote}We obtained our numerical solutions to Eqs.~(\ref{ode}) using normalized time, i.e., by solving the Schr\"{o}dinger evolution as ordinary differential equations in $\tau \equiv t\kappa$.  The associated normalized crystal length is therefore $\ell \equiv L\kappa/v$.   We then chose to discretize our crystal-length optimization in multiples of $10^{-3}v/\kappa$, to keep computation times within reasonable limits.
\end{thebibliography}
\end{document}